\documentclass[twocolumn,showpacs,preprintnumbers,amsmath,amssymb]{revtex4}
\usepackage{graphicx}
\usepackage{dcolumn}
\usepackage{bm}
\usepackage{txfonts}
\usepackage[ansinew]{inputenc}
\def\be{\begin{equation}}
\def\ee{\end{equation}}
\def\bee{\begin{eqnarray}}
\def\eee{\end{eqnarray}}
\usepackage{hyperref}
\hypersetup{
        colorlinks,
        linkcolor={blue},
        citecolor={blue},
       urlcolor={blue}
}

\begin{document}
\author{Giuseppina Modestino}
\email{modestino@lnf.infn.it}
\affiliation{%
INFN, Laboratori Nazionali di Frascati, I-00044, Frascati (Roma) Italy\\
}%


\date{\today}

\title{Multimessenger Research before GW170817
}

\date{\today }

\begin{abstract}
Linking the previous research that occurred over the last decades, I will try to provide some objective elements to evaluate the innovation of the joint observation of GW170817 and GRB 170817A and their occurrence detection, in light of preceding experiences regarding the experimental research of association between $\gamma$-ray bursts (GRBs) and gravitational waves (GWs).
Without debating about the phenomenological properties of astrophysical events, I propose a comparison between that result and the previous experimental research by the interferometer GW community, using a fundamental energy emission law, and including about fifteen years of accredited results regarding coincident detection.
From the present review, an intense and old pre-existing activity in the field of multimessenger observations emerges giving a first interesting fact. The widespread opinion that  joint detection of GW170817 and GRB 170817A has opened a new method in astrophysics does not find a robust reason. Moreover, some critical points highlight.
In the past, applying the same multimessenger method, numerous measures have been taken towards much brighter and much closer sources. Then, it would have been plausible to see joint signals even taking into account a worse sensitivity of the instruments of the time. At current time, there is only one event associated to a subthreshold GRB, compared to a long list of candidate events that would have been much more revealing. If these inconsistencies are admissible enough to lead to a claim, then the question arises about the interpretation of the long previous measurements carried out applying the same multimessenger observation method but without positive responses.\\
\end{abstract}

\maketitle

\section{Introduction}
Regarding  GW170817  by LIGO Scientific Collaboration and Virgo Collaboration,  and GRB 170817A by Fermi GRB Monitor, the joint detection (\href{ }{Abbott $et~al.$ \cite{abbott1,multi,grb_gw,siren,ligo_short}; Arcavi $et~al.$ \cite{nature2}; Goldstein $et~al.$ \cite{gold_l14}; Savchenko $et~al.$, \cite{integ_l15}; Kilpatrick $et~al.$, \cite{science}; Tanvir $et~al.$, \cite{kilo_l27}; Pan $et~al.$ \cite{sss17_l30}; Lipunov, $et~al.$, \cite{lipu}}) has been greeted as the opening of a new scientific era, the $multimessenger~astronomy$, meaning the search for GW signals and associated strong EM  emissions as GRBs are. The claim was accompanied by a particular emphasis underlying  novelty of the GW170817, such as to think that it is the result of a completely innovative observational method in the scientific panorama of GW research. 
Really, supernovae, magnetars and merger of binary compact systems like neutron star (NS) or black hole (BH) are always been considered sources for GW as well as for electromagnetic (EM) emission in a very wide frequency spectrum from highest $\gamma$ burst to X-ray, optical, infrared and soon on, so much that it is impossible to fully cite scientific references to justify this paradigm. (Anyway, from the ideological and historical point of view, in the introduction in \href{ }{Abbott $et~al.$ \cite{multi}}, a very large list of quotes can be found). It is widely believed that two time categories identify two types of progenitors, being long bursts ($> 2 s$) attributed to supernova explosions, while short GRBs to the fusion of two compact objects such as NS-NS or NS-BH. Although the bimodal time distribution has not been confirmed by all telescopes, as noticed by \href{ }{Tarnopolski \cite{tarno}}, this distinction is one of the main topics for the attribution of the GW170817 event to the merger of an NS-NS binary system. Even if any different progenitors may be in the case of black hole engine or millisecond-magnetar models for the production of GRBs, central engines may provide an unique theme between many classes of extremely luminous transient, from luminous supernovae to long and short GRBs (\href{ }{Levan \cite{grb_source}, Mereghetti \cite{mere2}, van Putten \cite{putten2}}). 
In that sense, dated back to the year 1968, the article \it"Prompt gamma rays and X rays from supernovae" \rm (\href{ }{Colgate \cite{colgate}}) can be reported as an historical example of interpretation as multifaceted phenomenon. GRBs were discovered between 1969 July and 1972 July using four widely separated $Vela$ spacecraft (\href{ }{Klebesadel \cite{vela}}). At first, the search was aimed to detect EM  fluxes near the times of appearance of supernovae. Subsequently, numerous mainly satellite telescopes were put into operation, with the precise scientific objective of detecting GRBs and interpreting their nature, and the experiments for the research of GWs have always referred to them, based on the fundamental hypothesis that GRBs and GWs have common origins (\href{ }{Fuller $\&$ Shi \cite{fuller}, Fryer $et~al.$ \cite{fryer}, van Putten \cite{putten1}}). 
The aim of this review is to examine the observations of GW170817 and GRB 170817A, taking into the account the preexisting  GW research investigating about the multiwavelength signals, trying to highlight as much as possible the well documented experimental activity related to the detection of EM transients understood as signatures extreme cosmological phenomena. 
 In the section \ref{gw_grb}, the initial GW experimental activity  aimed at GRB correlation research will be reported. That activity began at the end of the nineties of the past century, by the resonant bars (\href{ }{Coccia \cite{amaldi0}}) as detectors of GWs, able to monitor the Milky Way with the appropriate sensitivity according to the most accredited models (\href{ }{Misner $et~al.$ \cite{misner}}). Then, in Sect.\ref{ligo_grb}, the activity of GW interferometer detectors will be reported. The joint GRB and GW observations carried  out both using statistical technique analyzing until up thousand events (Sect.\ref{ligo_cum}), both dedicating special investigation to some special event or source (Sect.\ref{ligo_sgr}). In Sect.\ref{rer}, taking into consideration three fundamental aspects of multimessenger astronomy, the distance to the source, GW detectors sensitivity, and EM luminosity,  a model-independent term of comparison between the various observations will be defined.
In the Sect.\ref{discuss}, the comparison will be shown between the previous measurements and the GW170817 current evaluation, relatively applying  isotropic energy emission law \href{ }{(Abbott $et~al.$ \cite{ligo_short})} and the term just defined. 

 \section{Multimessenger astronomy and initial Gravitational Wave experiments}
 \label{gw_grb}
  The GW experiments started with \href{ }{Weber (\cite{virginia})}. 
 His controversial but ingenious activity has inspired many groups to undertake experimentation, especially using resonant detectors (\href{ }{Amaldi $et~al.$ \cite{amaldi1,amaldi2}; Astone $et~al.$ \cite{explorer,nautilus}; Mauceli $et~al.$ \cite{allegro}; Heng $et~al.$ \cite{austra}; Allen $et~al.$ \cite{igec}}). These kind of GW antennas produced an intense experimental activity (\href{ }{Pizzella \cite{pizzella}}) of which a significant part was dedicated to the search for correlations between astrophysical EM  transients and pulses from GW detectors. First of all there was a correlation study (\href{ }{Weber \& Radak \cite{radak}}) between  pulses recorded by an aluminium cylindrical bar between 1991 and 1992, and 80 GRB triggers from catalog of BATSE (Burst and Transient Source Experiment on NASA's Compton Gamma Ray Observatory-CGRO), a very revolutionary detector that was launched on 1991 (\href{ }{Paciesas $et~al.$ \cite{batse1}}). Aware that the hypotheses on the GRB were consistent with the collision models BH-NS and NS-NS, they presented the measurements performed with a bar of 3600 kg, showing compatibility with the gravitational radiation emitted by $1M_{\odot}$ at a distance of 1.5 Mpc. These measures have not been overly quoted by the scientific community, probably as result of the controversial issues raised by Weber that put into crisis the current model of GW detection at the end of the sixties, declaring a gravitational radiation evidence (\href{ }{ Weber \cite{weber69}}). In my opinion, beyond the specific merit, Weber's analysis of GRBs follows an irreproachable scientific method, mainly capturing the systematic nature of the astrophysical phenomenon, hence the reproducibility of observations and  with increasing in statistics also the possibility of improving the GW detector sensitivity. So much so that the procedure was taken up by several experimental groups later, operating both with resonant antennas (Sect.\ref{sbarre}) and laser interferometers (Sect.\ref{ligo_grb}). 
 Following, a summary of the best results from analyses aimed to establish statistical or especial associations with GRBs.\\
 \subsection{Resonant GW detectors}
\label{sbarre}
Basically, resonant detectors are metallic bars - typically aluminium - of mass $M$ and length $L$, that are lengthened by a quantity $\Delta L$ for a GW signal carrying an energy $E_s$. Computation of the GW strain amplitude h from the energy signal Es requires a model for the signal shape. Conventionally, a short pulse is considered with a flat spectrum on the resonance region and on equivalent bandwidth $\Delta \nu$. 
Based on the so called thermo-acoustic effect, the bar excitation mechanism is essentially described by the following formula (\href{ }{Pizzella \cite{pizzella97}, Astone \cite{nautilus2003}}) 
\be
\frac{\Delta L}{L}=h\approx \frac{L}{\tau_{g} v^2}\sqrt{\frac{E_s}{M}}
\label{acca}
\ee
where $v$ is the sound velocity in the specific metallic medium, and $h$ is the minimum discernible amplitude (SNR=1) for a GW signal with time duration $\tau_{g}\sim1/\Delta \nu$.
 Assuming $T_{n}$ (in kelvin units) as the temperature innovation for burst detection after optimum filtering for short signals, the signal energy is normally expressed as follows
\be
 E_s = k T_n
 \label{crio}
\ee
 where k is the Boltzmann constant. 
 Belonging to IGEC (International Gravitational Event Collaboration, \href{ }{Allen \cite{igec}, Astone $et~al.$ \cite{igec1,igec2}}), an international network of GW detectors, EXPLORER and NAUTILUS were long-lived devices as resonant bars suitable for GW experimental research, since for more than twenty, they represented reliable experimental apparata able to observe galactic GW signal generated by the conversion of about $0.001~M_{\odot}$. Having been made  by Roma group ROG (Ricerca Onde Gravitazionali), EXPLORER operated at CERN (Conseil Europ$\acute{e}$en pour la Recherche Nucl$\acute{e}$aire) since 1990, and NAUTILUS operated in LNF (Laboratori Nazionali di Frascati) of INFN (Istituto Nazionale di Fisica Nucleare, Italy) from 1995 until 2016, both using an aluminum bar of 2300 kg, a capacitive transducer
and a SQUID (Superconductive Quantum Interference Device) amplifier. Very similar to described devices there was also AURIGA (Antenna Ultracriogenica Risonante per l'indagine Gravitazionale Astronomica) (\href{ }{Prodi $et~al.$ \cite{auriga0}}), built by another Italian research group of INFN, and operating for a very long period at Legnaro (Padua, Italy). That were cryogenic detectors, so (as Eq.\ref{crio} can explain), the most sensitive GW detectors among resonant bars. In particular, NAUTILUS an AURIGA were the first ultra cryogenic detector being at $\sim 100mK$ for long time periods, reaching a typical noise temperature of a few thousandths of kelvin, with a sensitivity of $h_{min}\sim 10 ^{-19}$, representing the minimum GW amplitude detectable at SNR equal to 1 to GW signals over 1 Hz bandwidth around each one of the two resonant frequencies typically included in the interval $900-940~ Hz$. Both EXPLORER and NAUTILUS  were equipped with veto systems for cosmic-rays detection and amplitude calibration (\href{ }{Astone $et~al.$ \cite{cosmici,cosmici2}}) of the signal impinging the cryogenic bars. They have always represented the only GW detectors calibrated by a physical signal and not just by instrumental procedures. 
\subsubsection{Cumulative analyses}
\label{bar_stat}
The cumulative technique is very useful in multimessenger astronomy, allowing statistical tests and being particularly indicated for the search for correlations in the case of phenomena that present themselves in a systematic and recurring manner.
Data analysis techniques for measuring the effects of cosmic rays are very similar to those needed to search for correlations with any GRB signals. Precisely by developing algorithms for the maximization of the SNR in the detection of cosmic signals, and in the best adaptation to the real noise, analysis procedures have been implemented for the research of multiwavelength  correlations (\href{ }{Modestino $et~al.$ \cite{algo97,kolmo,algo_2000,cross}}). Generally, $N_\gamma$ data stretches from GW detectors are temporally aligned at $N_\gamma$ astrophysical events, then the eventual amplitude excess on the background fluctuations is exctracted. In principle, without taking into account the distribution of source distances, or aiming for fixed distance, the sensitivity should improve by $\sqrt{N_\gamma}$ factor, in consequence of the central limit theorem. For GW resonant detectors, thermal noise is reduced taking into account $T_{n}$ role as in the eq.\ref{acca}, the minimum amplitude improves in this way
\be
h_{N_{\gamma}}^2=\frac{h^2(T_n)}{\sqrt{N_\gamma}}.~~~~~~~~~~~~
\label{accag}
\ee
Several correlation analyses have been performed using that techniques:\\
$\bullet$ A study of the time correlation between GRB from BATSE and EXPLORER data was performed from April 1991 to December 1996 (\href{ }{ Astone $et~al.$ \cite{grb_explo}}). Examining five minutes around the time triggers,  1ms signals were excluded   with amplitude $h\geq 2.5~10^{-18}$.\\
$\bullet$ Selecting 20-minute output data intervals belonging to both EXPLORER and NAUTILUS data collected between 1991 and 1999 with individual stationary noise background of $T_n$=$12 mk$, 226 GRB triggers belonging that intervals was selected from the BATSE catalog.
Applying the cumulative procedure, and reducing the minimum detectable amplitude to $h=\ 2\times10^{-19}$, no time signature has been seen in a window of $20$ minutes around the GRB trigger times in the GW data background, with a confident level of 80\% (\href{ }{ Astone $et~al.$ \cite{search_mio_1998}}).\\
$\bullet$ By the same detectors, the analysis excluded the presence of a signal of amplitude $h \geq 5.4 \times 10^{-19}$, allowing a time delay between GW burst and GRB within $10 s $ (\href{ }{Astone $et~al.$ \cite{rog_2004}}).\\
$\bullet $ The previous result has been further improved to $h \geq 2.5 \times10^{-19}$ (\href{ }{Astone $et~al.$ \cite{cumulative}}), including 387 GRBs from BATSE 4B Catalog \href{ }{Paciesas $et~al.$ \cite{batse1,batse2}}) and observations of BeppoSax \href{ }{Boella $et~al.$ \cite{beppo1}, Frontera $et~al.$ \cite{beppo2}, Hurley $et~al.$ \cite{ipn_beppo}}), the Italian/Dutch satellite that estimate the first afterglow for a GRB (\href{ }{Costa $et~al.$ \cite{beppo3,beppo_x}}).\\
$\bullet$ During the year 2001, a data analysis was performed using again EXPLORER and NAUTILUS. Regarding to BeppoSax triggers and classifying them as $short$ $(< 5s)$ and $long$ $(\geq 5 s)$ GRBs, corresponding measurements $h_{short} = 2.1\times10^{-18}$ and $h_{long} = 2.0\times10^{-18}$ were obtained (\href{ }{Astone $et~al.$ \cite{beppo_rog}}). At that time, the result assumed a particular significance because BeppoSAX was the only GRB satellite in operation, as well as EXPLORER and NAUTILUS were the only GW detectors.\\
$\bullet$ At beginning of 2000s, using a single resonant GW detector,  AURIGA performed a search (\href{ }{Tricarico $et~al.$ \cite{auriga1}}) for an excess in coincidences with the arrival time of 120 GRB triggers collected in the BATSE catalogue between 1997 and 1998.\\
$\bullet$ Further, the data analysis group of  AURIGA extended the previous upper limit on the averaged GW energy released in $\pm 300 s$ around the GRB triggers, obtaining $h = 1.8 \times 10^{-18}$ at 95$\%$ confidence level (\href{ }{Tricarico $et~al.$ \cite{auriga2}}).\\

\subsubsection{Especial triggers and Soft Gamma Repeaters} 
\label{bar_sgr}
Since very consolidated models foresee the GW emission from them, among the most interesting and most focused sources are magnetars (\href{ }{de Freitas Pacheco \cite{freite}, Stella $et~al.$ \cite{stella1}}). 
Detected as persistent X-ray source at $\sim 10^{35} $ergs $ s^{-1}$, they are also called soft $\gamma$-ray repeaters (SGRs) because they occasionally emit energetic soft GRBs,  up to $\sim 10^{42} $ergs $ s^{-1}$, or even much more energetic events  (\href{ }{Thompson \& Duncan \cite{duncan1,duncan2}}). 
The large number of observed characteristics of SGRs including the bursting activity during the three giant flares detected to date (\href{ }{Mazets $et~al.$ \cite{maze1}, Helfand \& Long \cite{helfa}, Cline $et~al.$ \cite{cline}, Ioka \cite{ioka}, Mereghetti $et ~al.$ \cite{mere}, Hurley $et~al.$ \cite{hurley2}}) confirm their NS nature, and offer an effective evidence of the presence of very high magnetic field ($B\sim 10^{15}$G).
Such magnetars are typically found at galactic distances which makes these objects particularly interesting for the experimental studies between various astrophysical phenomenologies (\href{ }{Coccia $et~al.$ \cite{coccia}}). Following, performed by resonant bars, regarding SGR and other special triggers, several experimental results are reported.\\
$\bullet$ The GRB 980425 was the first one to be associated to a supernova occurring approximately the same time as  SN 1998bw (\href{ }{Soffitta $et~al.$ \cite{grb1998}, Tinney $et~al.$ \cite{sn1998_radio}, Woosley \cite{sn1998}}). Concerning that time, ROG group  presented the data analysis (\href{ }{Amati $et~al.$ \cite{superno}}) of the detector EXPLORER (with sensitivity  for a 1 ms pulse), using the trigger time from  BeppoSAX. The EXPLORER data exhibited no significant time signature around the GRB 980425. In spite of the low sensitivity, the measurements was important regarding active astrophysical observatories in coincidence at the time of the first evidence that GRBs and supernovae were related. \\
$\bullet$ During a particularly active phase of the SGR 1900+14 (\href{ }{Hurley $et~al.$ \cite{sgr1900}}) which took place on 28th August 1998, an analysis of coincidences between EXPLORER and NAUTILUS pulses was performed by \href{ }{Modena \& Pizzella (\cite{modena})} founding that a coincidence excess was concentrated during the period 7-17 September 1998 (21 on background of $ 8.60 \pm 0.09$), several days before the giant flare. \\
$\bullet $ The same previous data were correlated to the outburst of black-hole X-ray binary XTE J1550-564, an astrophysical object particularly suitable for multiwavelength observations (\href{ }{Wu $et~al.$ \cite{j1550}}). The correlation between the ROG data and the onset of the X-ray emissions in the energy range 20-100 keV was impressive but for explaining the experimental results, it should have been necessary to assume that the GW emitted power was significantly larger investigating the possible origin of the inferred signal (\href{ }{Drago $et~al.$ \cite{drago}}). Even if they refer to experimental data from another period, \href{ }{Coccia $et~al.$ (\cite{coccia})} were optimistic about the ability of the resonant bars to detect signals coming from the local galaxy or from the Virgo cluster, thinking of the SGRs as one of the most reliable sources.\\
$\bullet$ In the special framework of detectable sources, the SGR 1806-20 (\href{ }{Palmer $et~al.$ \cite{sgr1806}}) represents a very important scientific case being the most energetic explosion ever recorded for an astrophysical event. Studying  the giant flare occurring on December  $ 27^{th} $ 2004, the AURIGA group (\href{ }{Baggio $et~al.$ \cite{auriga_sgr}}) explored the frequency range 930-935 Hz, under the hypothesis of oscillating emission with a damping time of 100 ms. Expressing the result in terms of the dimensionless amplitude $h$, they found an upper limit of  $ 2.7\cdot 10^{-20}$, at the time of the hyperflare. \\
$\bullet$
The activity of the same SGR 1806-20 was studied (\href{ }{Modestino \& Pizzella \cite{mio}}) in correlation with the EXPLORER and NAUTILUS data, widening the observation to the bright outburst on October 5th 2004 as well as giant flare on December 27th (\href{ }{G\"{o}tz $et~al.$ \cite{gotz1}}). Two types of measurements were carried out. Averaging pulse amplitudes of GW detectors at time of outburst sequence and giant flare time, the presence of short pulses with energy $E_{gw}  \ge 1.8 \cdot 10^{49}$ erg was excluded with 90\% probability. Cross-correlating GW contemporary segments on 72 time flares occurring on October plus the giant burst on December, the corresponding probability distribution shows agreement with respect to a signal onset, with probability of chance result of about 1\%.

\section {Multimessenger observation by GW interferometers before GW170817}
\label{ligo_grb}
Search for correlation between GW bursts and GRBs has always characterised the activity of  interferometers since the beginning of their scientific runs, using the motivation according to which GRB progenitors are thought to be associated to hypernova explosions (\href{ }{Fruchter1 $et~al.$ \cite{fruchter1}, Hjorth $et~al.$ \cite{hjorth}})  or to coalescences of compact binary system (\href{ }{Eichler $et~al.$ \cite{eichler}}). 
To this aim, specific methods (\href{ }{Finn $et~al.$ \cite{finn1}, Finn \& Mohanty \cite{finn2}, Kalmus $et~al.$ \cite{kalmo}}) have been implemented, and several measurements have been performed above all targeting short-duration signals near the time of the EM triggers relating upper limits on the GW energy radiated $E_{gw}$. In general, for the expected amplitude, an upper limit  can hardly relate to the energy radiated by the source,  without assuming a model that requires $a~priori$ waveform definition for data analysis. But treating emissions at high frequencies ($\sim$1kHz), as the explosive stage of supernovae and flares emitted by magnetars, it is not always possible to adopt this procedure because the emissive dynamics are not clear unlike inspiral binary system, continuous waves or stochastic background. Thus, many choices are possible among models to be adopted and the results of analyses can differentiate themselves very much even aiming at the same physical event. These include astrophysical waveforms, such as the phenomenological simulations of emission from typical source like supernovae, as well as ad hoc waveforms such as Gaussians, damped sinusoid and sine Gaussians (\href{ }{Abbott $et~al.$ \cite{ligo_s1}}). The comparison between different measurements is also complex no less than they have been performed with the same instrumentation. But trying to synthesize the previous multimessenger activity in a unique experimental framework, it is useful to refer to the formula that generally expresses the GW energy emitted isotropically (\href{ }{Abbott $et~al.$ \cite{ligo_short}, Sutton \cite{sutton}}), limiting in a narrowband with central frequency $f$:
\be
E_{gw}\simeq\frac{\pi^2c^3}{G}D^2f^2h_{rss}^2,
\label{energia}
\ee
where $D$ is the distance between the emitting source and the observer, and $h_{rss}$ is defined as the root sum-squared strain amplitude of GW signals impinging the detector  
\be
h_{rss}^2={\int_{-\infty}^{\infty}[h_+^2(t)+h_{\times}^2(t)]dt},
\label{h_rss}
\ee
with the two quadrupolar  components $h_+$ and $h_\times$.\\
Following, a list of works performed by LIGO, Virgo and GEO600, since 2003 up to GW170817, including the scientific runs S2-S6, O1-O2 for LIGO and Advanced LIGO, VS1, VS2, VS3 for Virgo.
A synthesis can be found also in the Table \ref{inter_cum} and \ref{inter_sing}. 
\begin{table*}
\caption{
The table shows eight studies (see Sect.\ref{ligo_cum}) performed by statistical approach that could be sensitive to the cumulative effects of any weak GW signal. The GRBs were supplied by telescopes belonging to the list of Table\ref{cara_stru}. In the 1st column there is the number of the GRB triggers used in the analyses. Then, the relative years and GW detectors with specific runs are shown. In the last column, there is the median value of the distances between the source and the detector GW. Most of them represent the limit value calculated assuming an emission of energy equal to 0.01 $M_{\odot}$ c$^2$, in the best range of frequency sensitivity spectrum, typically 150-500 Hz.
}
\begin{tabular}{|c|c|c|c|c|}
\hline
&&&&\\
\# GRB Triggers & Observation Years &GW Detector &Run& Distance\\
& &(Measurement)& &[Mpc]\\
\hline
&&&&\\
39 &2003-2005& LIGO&S2 S3 S4&\\
&&(Abbott $et~al.$ \cite{ligo_39grb})&&\\
\hline
&&&&\\
137 &2005-2007& LIGO \&Virgo& S5 VSR1&12\\
&&(Abbott $et~al.$ \cite{ligo_137grb})&&\\
\hline
&&&&\\
22 &2005-2007&LIGO&S5 VSR1& 6.7\\
&&(Abadie $et~al.$ \cite{ligo_22grb})&&\\
\hline
&&&&\\
50&2006-2007&LIGO&S5&33\\
&&(Aasi $et~al.$ \cite{ligo_long_grb})&&\\
\hline
&&&&\\
223  &2005-2010&LIGO \& Virgo&S5 S6 VSR1 VSR2 VSR3&13\\
&&(Aasi $et~al.$ \cite{ligo_223grb})&&\\
\hline
&&&&\\
154&2009-2010&LIGO \& Virgo&S6 VSR2 VSR3&17\\
&&(Abadie $et~al.$ \cite{ligo_virgo154grb})&&\\
\hline
&&&&\\
129&2006-2011&LIGO Virgo \& GEO600&S5 S6 VSR1 VSR2 VSR3&0.3\\
&&(Aasi $et~al.$ \cite{lvgc_129})&&\\
\hline
&&&&\\
41&2015-2016&Advanced LIGO&O1&71\\
&&(Abbott $et~al.$ \cite{ligo_grb15})&&\\
\hline
\end{tabular}
\label{inter_cum}
\end{table*}

\begin{table*}
\caption{Multimessenger observations (see Sect.\ref{ligo_sgr}) conducted by GW interferometers towards candidate sources for NS binary merger, and towards SGRs located in the local galaxy. The strain sensitivity of the GW interferometer is referred to run best value at 1kHz, in the frequency spectrum, and it is used for the calculation of the comparative parameter RER described by Eq.\ref{rela_expe}. The joint observation GW170817 and GRB 170817 is included and is taken as reference value $GRB_0$ in the same Eq.\ref{rela_expe}. 
}
\begin{tabular}{|c|c|c|c|c|c|}
\hline
&&&&&\\
Triggers & Observation Year&Distance &Fluence&Strain sensitivity $(\sim 1kHz)$&RER $(\sim 1kHz)$\\
(Measurement)& &[Mpc]&[ergs/cm$^2$]& $[\sqrt{1/Hz}]$&\\
&&&&&\\
\hline
&&&&&\\
GRB 030329&2003&800 &$\sim 10^{-4}$&$2\times 10^{-21}$&$1.2\times10^{-2}$\\

(Abbott $et~al.$ \cite{ligo_grb03})&&&&&\\
\hline
&&&&&\\
QPO from SGR1806-20 &2004&0.01&$5\times10^{-3}$&$4\times10^{-22}$&$2.6\times 10^{4}$\\

(Abbott $et~al.$ \cite{ligo_sgr1})&&&&&\\
\hline
&&&&&\\
 SGR1806-20  (Giant Flare)&2004&0.01&2&$4 \times 10^{-22}$&$5.2\times 10^{5}$\\
SGR 1900+14 (190 GRBs) &2005-2006&&&$3 \times10^{-22}$&$3.7\times 10^{3}$\\
&&&&&\\
(Abbott $et~al.$ \cite{ligo_sgr2})&&&&&\\
\hline
&&&&&\\
GRB 051103&2005&3.6&$2.3\times 10^{-5}$&$3 \times10^{-22}$&6.5\\

(Abadie $et~al.$ \cite{ligo_grb05})&&&&&\\
\hline
&&&&&\\
SGR 1900+14 (storm) &2006&0.01&$10^{-4}$&3$ \times 10^{-22}$&$3.7\times 10^{3}$\\

(Abbott $et~al.$\cite{ligo_sgr3})&&&&&\\
\hline
&&&&&\\
1279 GRBs from&&&&& \\
&2006-2009&&&&\\
SGR 0418+5729 &&0.002&&&1.1$\times 10^{4}$\\
SGR 0501+4516 &&0.001&&&2.2$\times 10^{4}$\\
AXP 1E 1547.0-5408 &&0.004&$\sim 2 \times 10{-5}$&$3 \times 10^{-22}$&$5.4\times 10^{3}$\\
SGR 1627-41 &&0.01&&&2.2$\times 10^{3}$\\
SGR 1806-20 &&0.01&&&2.2$\times 10^{3}$\\
SGR 1999+14 && 0.01&&&2.2$\times 10^{3}$\\
&&&&&\\
(Abbott $et~al.$ \cite{ligo_sgr4})&&&&&\\
\hline
&&&&&\\
GRB 070201& 2007&0.77&$1.6\times 10^{-5}$& $5 \times 10^{-22}$ &15.2\\

(Abbott $et~al.$ \cite{ligo_grb07})&&&&&\\
\hline
&&&&&\\
GRB 150906b&2015&54 &$ 2.8\times10^{-5}$& $2 \times 10^{-23}$&7.4\\
(Abbott $et~al.$ \cite{ligo_grb15}) &&&&&\\
\hline
&&&&&\\
GW170817&2017&40 &$1\div 3 \times 10^{-7}$&$2\times10^{-23}$&1\\
&&&&&\\
(Abbott $et~al.$ \cite{abbott1})&&&&&\\
\hline
\end{tabular}
\label{inter_sing}
\end{table*}
\subsection{Cumulative analyses by interferometers}
\label{ligo_cum}
By GW interferometer detectors, several experimental studies have performed by statistical approach that could be sensitive to the cumulative effects of any weak GW signal. GRB triggers are mainly from satellites belonging to the InterPlanetary Network (IPN)[\cite{ipn0}], a coordination running telescopes detecting GRBs and optimising their reception. Typically, the method consists in combining (\href{ }{Finn \& Mohanty \cite{finn2}, Kalmus $et~al.$ \cite{kalmo}}) output of two or more GW detectors, then evaluating the statistical difference between output at times of GRB ($on-source$) and output at other times ($off-source$) not associated with GRB would reveal clearly a signature if a GW-GRB association exists.\\
$\bullet$ Starting from 2003, LIGO collaboration set upper limits (\href{ }{Abbott $et~al.$ \cite{ligo_39grb}}) on the strain amplitude $h_{rss}$ of sine-gaussian waveforms at the times of 39 GRB triggers provided by the IPN including Konus-W (\href{ }{Pal'shin $et~al.$ \cite{ipn_konus}}), HETE-2 (\href{ }{Hurley $et~al.$ \cite{ipn_hete}}), INTEGRAL (\href{ }{Winkler $et~al.$ \cite{integ1}}), and Swift (\href{ }{Lien $et~al.$ \cite{swift2}}), and  distributed via the GRB Coordinates Network (GCN)[\cite{gcn_home}]. The GRB triggers were during LIGO science runs S2, S3 and S4, with the best sensitivity of the order $\sim 10^{-22}\sqrt{1/Hz}$, between 100-1000 Hz.\\
$\bullet$ LIGO \& Virgo collaborations presented the result (\href{ }{Abbott $et~al.$ \cite{ligo_137grb}}) of a search for GW bursts associated with 137 GRBs collected during the 5th LIGO science run and 1st Virgo science run, between 2005-2007. Assuming isotropic emission at median distance of 12 Mpc, they placed lower bounds on 0.01 M$_\odot$ c$^2$ for GW energy at 150 Hz.\\
$\bullet$ During the same science runs, LIGO and Virgo collaborations presented a search (\href{ }{Abadie $et~al.$ \cite{ligo_22grb}}) for coincident signature for 22 GRBs provided by GCN. Examining a few seconds around the trigger time of the GRBs, they excluded the merger of NS-NS and NS-BH at median distance 3.3 Mpc and 6.7 Mpc correspondingly. The analysis was interesting including special bursts as GRB 070201 (\href{ }{Golenetskii $et~al.$ \cite{070201}}), spatially localised into M31, a galaxy less the 1 Mpc from Earth. \\
$\bullet$ Then the same collaborations  investigated (\href{ }{Aasi $et~al.$ \cite{ligo_long_grb}}) models of long-lived ($\sim$ 10-1000s) GRBs typically associated to the extreme core collapse of massive star and its protoneutron remnant, and so to the GW emission. They used data from LIGO's fifth science run and 50 GRB triggers from the Swift experiment, finding no evidence of long-lived GW transients and setting 90\%\rm C.L. upper limits on the GW emission ranging three orders of magnitude for the GW fluence, depending on the three waveform adopted models. \\
$\bullet$ A search for GWs  was performed (\href{ }{Aasi $et~al.$ \cite{ligo_223grb}}) including 223 GRBs detected by the IPN during 2005-2010. 
The interferometer collaborations LIGO \& Virgo placed lower bounds on the distance to the source, finding a median of 13 Mpc, assuming a gravitational-wave emission energy of $10^{-2} M_{\odot}$ at 150 Hz. For the 27 short-hard GRBs, 90\% confidence exclusion was fixed for the distances to two source models: a binary NS coalescence, with a median distance of 12 Mpc, or the coalescence of  NS-BH, with a median distance of 22 Mpc. \\
$\bullet$ Searching  for coalescence of compact binary system during the same science runs S3-S4, although LIGO
detectors could probe to distances as far as tens of Mpc, no GW signals were identified in
the 1364 hours of data (\href{ }{Abbott $et~al.$ \cite{binary08}}).\\
$\bullet$ Referring to S6 run and VS2-VS3 runs, the same LIGO and Virgo presented the results of a search for association with 154 GRB that were detected in 2009-2010 (\href{ }{Abadie $et~al.$ \cite{ligo_virgo154grb}}). The GW energy emission was excluded for sources at 17 Mpc and at frequency 150 Hz. From the whole sample, short-hard GRBs were extracted in order to evaluate eventually the presence of NS-NS or BH-NS mergers. That were excluded taking in the account distances lower than 16 and 28 Mpc respectively.\\
$\bullet$ Using data between 2006-2011, the analysis(\href{ }{Aasi $et~al.$ \cite{lvgc_129}}) was performed by LIGO, Virgo and GEO600, correlating to 129 GRB triggers. The result was no evident signature either with any individual GRB or with the whole sample, placing lower bounds on the distance of 0.3 Mpc for GW emission energy of $0.01 M_{\odot}c^2$, at 1kHz.\\
$\bullet$  The sensitivities during LIGO and Virgo joint science runs in 2009-2010 were also compared (\href{ }{Aasi $et~al.$ \cite{ligo_kilo}}) to several model light curves from possible sources of interest, and  imminent prospects were discussed for joint GW-optical observations of this type.\\
$\bullet$  LIGO, Virgo and IPN collaborations found no evidence (\href{ }{Abbott $et~al.$ \cite{ligo_grb15}}) of a GW signal for any of the 41 GRBs detected during the first observing run of the Advanced LIGO. They used the assumption that GWs were emitted with an energy of $10^{-2}~ M_\odot c^2$, within the 16-500 Hz band. In the same study, they put several upper limits for the whole data set and also for selected subsets  for different coalescence system configurations at various distances of the order of ~ 100 Mpc. They also discussed the results of the search for GWs associated with GRB 150906B in detail, excluding a binary system as the progenitor of the event with confidence $>99\%$, in NGC 3313, a local galaxy at a luminosity distance of 54~Mpc from~Earth.

\subsection{Especial triggers and SGR observation by GW interferometers}
\label{ligo_sgr}
Despite their unpredictable behaviour, the observational properties of SGRs and their interpretation principally in the magnetar model (\href{ }{Mereghetti \cite{mere2}}) give the perspective that these objects were among the most promising sources of GWs (\href{ }{Owen \cite{owen}, Horvath \cite{horv}, Horowitz \& Kadau \cite{horo}}). In this direction, well-established studies exist (\href{ }{Ioka \cite{ioka}, Corsi \& Owen \cite{corsi}}) where it has been demonstrated that changes in the hydromagnetic deformation of a magnetar can provide an energy reservoir comparable to LIGO and Virgo sensitivity for some time, considering also that the proximity ($\sim10$ kpc) makes these objects intriguing for GW searches, enhancing chances of detectability. To follow, several experimental studies carried out by interferometers coinciding with SGR emissions and other special $\gamma$-triggers.\\
$\bullet$ In one of its first experimental studies on the association between GW and  EM  emission, LIGO collaboration reports frequency-dependent upper limit (\href{ }{Abbott $et~al.$ \cite{ligo_grb03}}) on the strength of the gravitational waves associated with GRB 030329 (\href{ }{Stanek $et~al.$ \cite{spn2003}}), the burst that has definitively demonstrated the connection between gamma emission and supernova. The strain sensitivity for optimally polarised bursts was less then $h_{rss} =6 \times10^{-21}\sqrt{1/ Hz}$ around 250 Hz, the most sensitive region.\\
\begin{figure*}
\includegraphics[width=0.9 \linewidth]{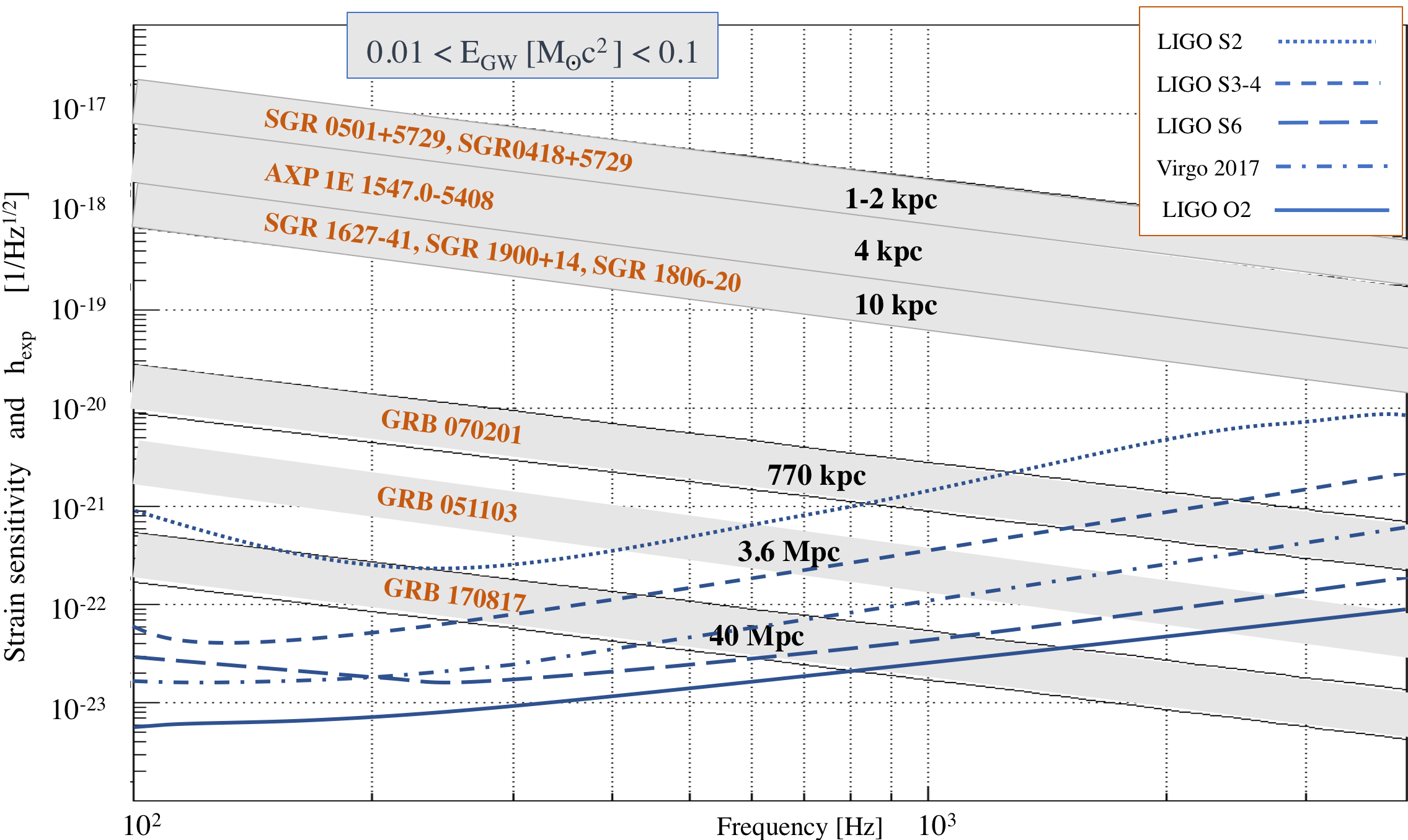}
\caption{
Strain sensitivity of the GW interferometer detectors since 2003 up to 2017. From top, dotted line, dashed line, larger dashed line and continuous line are respectively the sensitivities during LIGO scientific runs S2, S3-4, S6, and O2 observing run of Advanced LIGO. 
The dashed-dotted line is Virgo strain sensitivity at the time of the GW170817. The grey bands represent the expected region of amplitude $h_{exp}$, for a narrowband GW signal with fixed energy content $E_{gw}$, at different source distances. The Eq.\ref{energia} is used under the very credible assumption that only a little fraction of the total energy is emitted in the form of GWs (\href{ }{Abbott $et~al.$ \cite{ligo_short}}). Considering a system of a few solar mass, a range of values between 0.01 and 0.1 is evaluated. The distances are fixed at the known $\gamma$ burst sources used by GW interferometers for several correlation experimental measurements (see Table\ref{inter_sing}).
\label{hrss_spectrum} }
\end{figure*}
\begin{figure}
\includegraphics[width=1\linewidth]{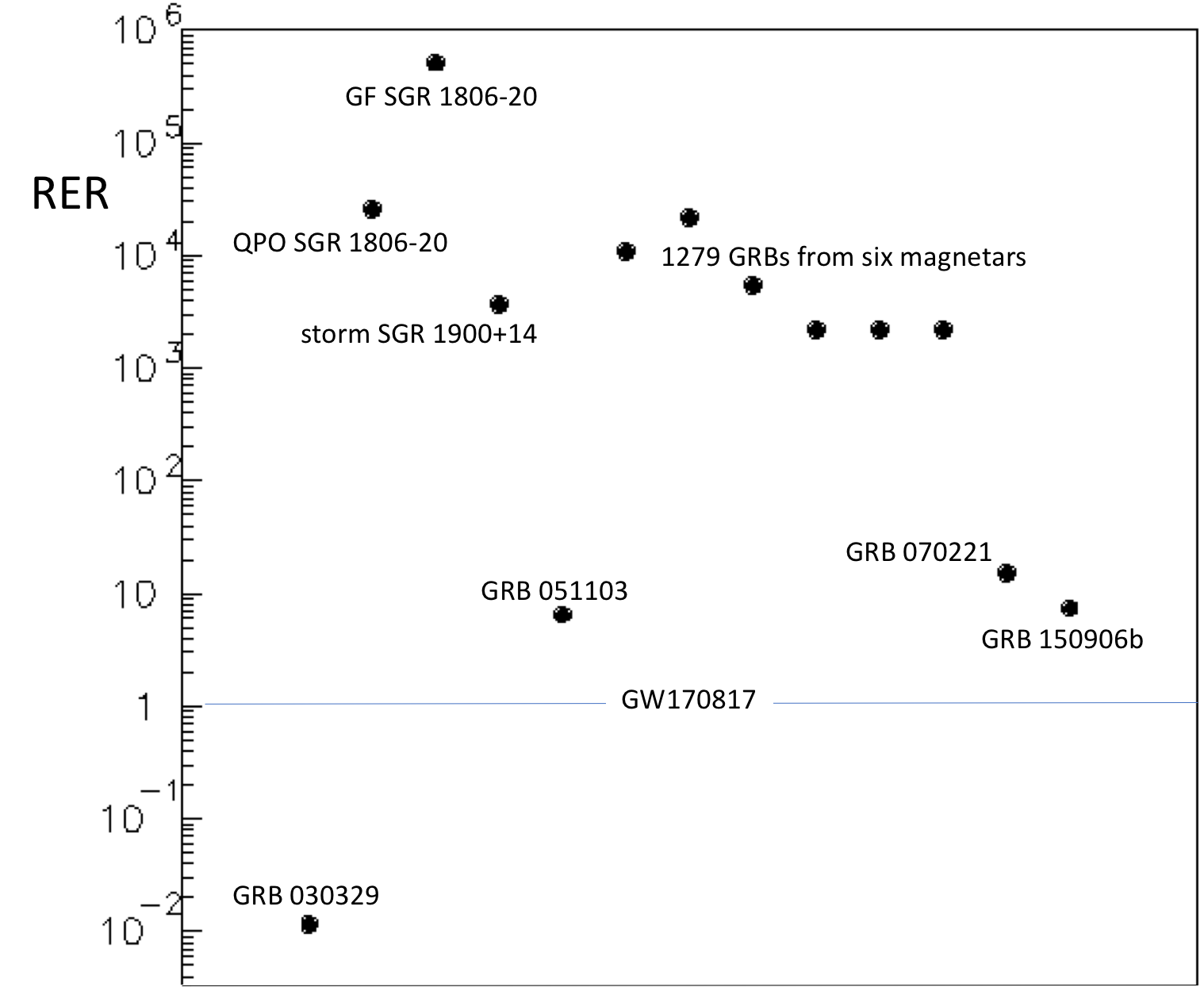}
\caption{Relative expectation rate (RER) on several multimessenger measurements performed by the GW interformeters since 2003. Is has been evaluated using the definition \ref{rela_expe}, extracting the physical parameters from the observations reported in the Sect.\ref{ligo_sgr} and in the Table \ref{inter_sing}, and fixing GRB 170817A as reference trigger. As it can easily noted, excepting for the first GRB 030329, any other value is considerably higher, even by several orders of magnitude.
\label{fig_re} }
\end{figure}

$\bullet$ 
In relation to the special outburst on December $27^{th}$ from SGR 1806-20, the LIGO collaboration (\href{ }{Abbott $et~al.$ \cite{ligo_sgr1}}) examined the pulsating tail of the burst which revealed the presence of quasi-periodic oscillations (QPOs) in the X-ray light curve, as RXTE (\href{ }{Israel $et~al.$ \cite{israel}}) and RHESSI satellite (\href{ }{Watts \& Strohmayer \cite{watts}}) had detected. LIGO found no excess and they set several upper limit levels on the GW emitted energy, racing from $\sim 10^{47}$ ergs, depending on time and frequency radiation. \\
$\bullet$ Further, the same LIGO collaboration presented the results of short-duration GW events associated with SGR 1806-20 and SGR 1900+14 storms  occurred during the year starting from November 2005 (\href{ }{Abbott $et~al.$ \cite{ligo_sgr2,ligo_sgr3}}). Including the giant flare, they analysed almost two hundred  events finding no evidence of any association. Depending on simulation types frequency ranges, several upper limit were estimated, referring to many orders  of magnitude (from $10^{45}$ erg to almost $10^{53}$ erg) for hypothetical isotropic GW energy emission.\\
$\bullet$ Virgo carried out the first coincident analysis (\href{ }{Acernese $et~al.$ \cite{virgo1}}) between its data and a GRB at time of GRB 050915a (\href{ }{Grupe $et~al.$ \cite{050915}}), during C7, one of its science runs.\\
$\bullet$ A coincidence examination has been carried out by LIGO (\href{ }{Abadie $et~al.$ \cite{ligo_grb05}}) with respect to event GRB 051103 (\href{ }{Golenetskii $et~al.$ \cite{051103}}) which energy release was  $\approx 4.5\times10^{46}$ erg, assuming an isotropic emission from a source in M81 (D = 3.6 Mpc). Expecting astrophysical correlations from binary coalescence, a few seconds on source window (-5,+1) was chosen around the  EM  prompt, but no GW signature emerged with the respect to the results from the off-source evaluation. Then, they concluded that it was highly unlikely that the progenitor for GRB 051103 was a compact binary merger in M81, but it would be indeed one of the most distant SGR giant flares observed to date.\\
$\bullet$ In a subsequent study (\href{ }{Abadie $et~al.$ \cite{ligo_sgr4}}), LIGO extended the collaboration for analyzing  data also from Virgo, and GEO600 detectors relatively a sample of 1279 EM triggers  from six SGR emitters occurring between November 2006 and June 2009. The magnetars were  thought very close from Earth,  SGR 0501+4516 and SGR 0418+5729 at about 1 kpc; AXP 1E 1547.0-5408 also known as SGR 1550-5418 is at 4 kpc, and SGR 1627-41 from which 54 peaks had been analysed selecting from Swift light curves. In the same analysis, the two most famous magnetars SGR 1806-20 and SGR 1900+14 can be found. No evidence was found for GW emission.\\
$\bullet $ Looking for signals associated with GRB 070201, no plausible GW signals were identified cross-correlating  on time data from the LIGO H1 and H2 detectors (\href{ }{Abbott $et~al.$ \cite{ligo_grb07}}).\\

\subsection{Relative Expectation Rate}
\label{rer}
  To give a synthetic and comparative idea in a multimessenger experimental framework, the relative expectation rate (RER) is defined in terms of the ratio between the amplitude of the GW wave expected for the isotropic emission of energy associated with the n-th $GRB_n$, and the expected amplitude for the reference $GRB_0$. It involves three parameters, distance between the source and observer, EM fluence of GRB and  GW detector sensitivity expressed in terms of strain amplitude $h_n(f)$ and $h_0(f)$ revealed in the frequency domain. Taking into the account the proportionality between total isotropic energy $E_{iso}$ describing  the energy emitted by GRBs, and the relative measured EM fluence ($E_{iso} \sim Fl$), practically:
\be
RER(f)\equiv \frac{D_0}{D_n} ~\sqrt{\frac{Fl_n}{Fl_0}} ~\frac{h_0(f)}{h_n(f)},
\label{rela_expe}
\ee
where $D_n,~Fl_n$ and $D_0,~Fl_0$ are the corresponding distances and the measured fluence from the sources of $GRB_n$ and $GRB_0$. For the cases reported in the previous paragraph, that parameter is calculated considering the n-th single GRB as from the Table\ref{inter_sing}, and the event GRB$_0$ as GRB 170817A. The results are reported in the last column of the Table\ref{inter_sing}, and in the Fig.\ref{fig_re}. 

\section{Measurement Comparison and Discussion}
\label{discuss}
 In Fig.\ref{hrss_spectrum}, the expected amplitude $h_{exp}$ is shown as a 
 function of emission frequency, calculating it from the Eq.\ref{energia}, extracting the physical parameters from the most significant EM  triggers, and from accredited models which see the systems constituted by some solar mass as the most widespread. Referring to a very common assumptions, the range $0.01 \leq E_{gw}\leq 0.1$ can be adopted, as done in the six grey bands of the Fig.\ref{hrss_spectrum}, each for the most representative $\gamma$-triggers at relative distances. 
The upper three bands concern the expected $h_{exp}$ values for the six galactic magnetars detected at distance $1\leq D \leq 10$ kpc, then we find GRB 070201 at $D=770$ kpc, GRB 051103 at $D=3.6$ Mpc and GRB 170817A identified as GW170817 at $40$ Mpc. In the same figure, five curves are plotted indicating the best sensitivity of the GW interferometer detectors during several years, proceeding from 2003 until 2017. Essentially, they correspond to LIGO best sensitivity during the scientific runs from S2 to S6, during the second observing run (O2) of Advanced LIGO, and during Virgo science run on 2017. As shown, excluding the recent GRB 170817A, most of the observed sources would have been within reach of the instruments being higher than the relative sensitivity curves, at least in the best frequency range around 150 Hz. A summary evaluation and comparison of the measures are provided also by the RER parameter defined in Eq.\ref{rela_expe} and plotted in Fig.\ref{fig_re}. It can be easily noted that for each $GRB_n$, the corresponding expectation value is much higher then 1, even more than several magnitude orders, so indicating in any case that a GW signal detection would be much more probable in the past than for the recent GW170817. 
Being GRB 170817A two to six orders of magnitude less energetic than other short GRB (\href{ }{Abbott $et~al.$ \cite{multi}}), in light of theoretical NS merger models and existing GRB classification, it can hardly fit into a simple phenomenological scheme, as noted also by (\href{ }{Horv\'{a}th $et~al.$ \cite{horva}}). 
From a single event, even sub-threshold, it is difficult to deduce a systematic effect in order to objectively identify a theoretical model and so assign an undeniable progenitor. 
About fifty years ago, the scientific GW community debated about such a case where divergences between the theoretical model and experimental results were found, and the opinion of the most eminent scientific personalities was to wait for further validation before claiming the discovery. In order to frame the episodes in a scientifically correct manner, the words  of \href{ }{Press \& Thorne (\cite{press_thorne})}  are useful: 
\it"...It is characteristic of important scientific puzzle that before the solution is known all possibilities look equally implausible..."\rm. In this way they commented Weber's experimental observations (\href{ }{Weber \cite{weber69}}), noting the absence of theoretical models to suitably fit the related results. But feeling that the experimental developments were close in order to confirm or confute certain results in terms of gravitational waves, the authors considered it appropriate to wait a few more years (at the latest fifteen) to improve the right detection technology and thus perform reliable measurements. Unfortunately they were very optimistic and the years were not so few, but in the meantime considerable progress has been made so that the interferometric community has recently declared several BH merger detections (\href{ }{Abbott $et~al.$ \cite{bh1,bh2,bh4,bh3}}). This ability should make the GW community confident of its competence to systematically detect mergers of compact binary systems, and associate them with GRBs. So, having more available statistics would be easier to interpret the correlations, setting constraints on known models or reformulating ones.
 \section{Conclusions}
\label{conclu}
As reported, in the past there has been an intense experimental activity aimed to multimessenger physics. The well-documented experimental experience makes rather inappropriate the term $unprecedented$ if it is related to the 
multiwavelength  typology of GW170817 and GRB 170817A joint detection. At the same time, several scientific questions arise concerning both the measures in the past and the recent joint survey.  But, if the GW community is so confident that there is unambiguous  correlation despite being with a sub-luminal GRB, and if a new era has really opened up, then the same expertise group should also explain how to close the previous era, when the numerous measurements performed and compared to much closer and energetic sources have not given expected results.\\

\acknowledgements
I thank Giovanni Mazzitelli for his encouragement and suggestions. I gratefully acknowledge the friendliness of collaborating of Francesco Ronga and researchers of former ROG group. I also acknowledge fruitful discussions with colleagues of the Research Division of $Laboratori ~Nazionali ~di ~Frascati ~of ~INFN.$

\end{document}